\documentclass[12pt]{emulateapj}
\usepackage{natbib}
\usepackage{color}
\tightenlines

\newcommand     \ch    {{\rm ch}}
\def     \bv    {{\bf v}}
\def     \br    {{\bf r}}
\def     \bFr    {{\bf F}}

\def     \kin   {{\rm kin}}
\newcommand     \cm    {{\rm cm}}  
\newcommand     \s       {{\rm s}}

\newcommand     \K       {{\rm K}}
\def     \H   {{\rm H}}

\def     \cf    {{\rm cf}}
\def     \gas  {{\rm gas}}

\def    \tot      {{\rm tot}}

\def    \drag   {{\rm drag}}
\def    \B		  {{\rm B}}
\def    \d        {{\rm d}}

\newcommand     \Zmean {$\langle Z\rangle$}
\def   \chrg    {{\rm chrg}}

\def	\Angstrom	{\,{\rm \AA}}		

\def \bea {\begin{eqnarray}}
\def \ena {\end{eqnarray}}

\begin{document}
\title{Acceleration of Small Dust Grains due
to Charge Fluctuations}
\author{Thiem Hoang \& A. Lazarian}
\affiliation{
Department of Astronomy, University of Wisconsin, Madison, WI 53706, USA}

\begin{abstract}
We consider the acceleration of very small dust grains including Polycyclic
Aromatic Hydrocarbons (PAHs) arising from the electrostatic interactions of
dust grains that have charge fluctuations in time due to charging events.
We simulate the charge fluctuations of very small grains due to
their sticking collisions with electrons and ions in plasma, and the emission of
photoelectrons by UV photons using Monte Carlo method. We identify
the acceleration induced by the charge fluctuations as the dominant
acceleration mechanism of very small grains in the diffuse interstellar
medium (ISM). We show that this acceleration mechanism is efficient
for environments with low degree of ionization (i.e., large Debye length), where the charge fluctuations are
slow but have a large amplitude. We also discuss the implications of
the present mechanism for grain coagulation and shattering in the diffuse ISM,
molecular clouds, and protoplanetary disks.
\end{abstract}

\maketitle
\section{Introduction}

Dust is an important constituent of the interstellar medium (ISM), molecular clouds,
and accretion disks (see Whittet 2003; Draine 2009, 2011). It gets involved in many
key processes, for instance, it controls heating and cooling of the ISM (see Draine
2003; Tielens 2005), reveals magnetic fields through grain alignment
(see Lazarian 2007 for a review), and interferes with attempts to measure
the temperature anisotropy and polarization of the cosmic microwave background (CMB)
radiation (see Lazarian \& Finkbeiner 2003; Fraisse et al. 2009; Dunkley et al. 2009).

Very small dust grains of size $a\le 100\Angstrom$ with a
notable fraction of polycyclic aromatic hydrocarbons (PAHs, hereafter
very small grains and PAHs are used interchangeably) radiate
electric dipole emission, which is an important component of CMB foregrounds
(Draine \& Lazarian 1998; Hoang et al. 2010, 2011). PAHs
in protoplanetary disks (PPDs) around young intermediate mass (Herbig Ae/Be)
stars (Acke \& van den Ancker 2004) and in outer layers of low mass (T-Tauris) stars
(Geers et al. 2006; Oliveira et al. 2010) may affect the magnetorotational
instability (MRI) due to their influence on the environment ionization
(Bai 2011; Perez-Becker \& Chiang 2011).

Most properties of dust, including light extinction, electron photoemission,
and chemical activity depend not only on grain chemical composition, but also
on their sizes. In astrophysical environments, grain size is affected by grain-grain
collisions, which depends on their relative velocities.
The minimal velocity of grains corresponds to their thermal velocity
at the ambient gas temperature arising from Brownian motions. These velocities
are usually assumed in the models of dust coagulation (Ossenkopf 1993;
Dominik \& Dullemond 2008). It has been long known that large scale
hydrodynamic motions associated with turbulence can make grains
move faster (see Draine 1985), but detailed calculations applicable to
astrophysical environments started to appear only a few years ago.

Since most astrophysical media are magnetized and grains are charged, the
hydrodynamic treatment of acceleration is not adequate. A proper
treatment of the grain acceleration through the resonant interactions of charged grains with
magnetohydrodynamic (MHD) turbulence has been developed relatively recently
(Lazarian \& Yan 2002; Yan \& Lazarian 2003; Yan et al. 2004; Yan 2009; Hoang et al. 2011b).
This treatment makes an extensive use of the advances in the theory/numerical studies of
compressible MHD turbulence (see Cho \& Lazarian 2003; Kowal \& Lazarian 2010 and ref. therein) and provides the mathematical 
formalism of the second-order
Fermi acceleration of charged grains interacting with MHD turbulence.

The acceleration due to the resonant interactions of MHD turbulence
with charged grains decreases with decreasing grain size. Such decrease arises from the
fact that the Larmor radius of charged grains becomes smaller as the grain
mass decreases, and correspondingly, grains have to interact with smaller
and less powerful turbulent fluctuations.
In addition, compressible fluctuations (i.e., fast modes), which were identified in
Yan \& Lazarian (2003) as the most efficient source of acceleration, get
suppressed at small scales due to plasma viscous damping, while
the Alfv\'{e}nic mode gets inefficient for acceleration at the small scales due to
anisotropy (see Yan \& Lazarian 2003 for more discussion). The resonant acceleration
for grains with sizes $\leq 10^{-5}$~cm becomes rather inefficient in most media
discussed in Yan \& Lazarian (2003) and Yan et al. (2004, hereafter YLD04).
These conclusions were also confirmed by Hoang et al. (2011b) who considered
the non-linear effects of resonant acceleration (Yan \& Lazarian 2008) and
the transit time damping acceleration (TTD) of charged grains.

A new mechanism of astrophysical grain acceleration was sketched in Ivlev et al. (2010),
where rough estimates of grain velocities arising from charge fluctuations
in the ISM were provided. The study employed Fokker-Planck equation and
assumed that (1) charge fluctuations are fast, i.e., the relaxation timescale
for charge fluctuations is much shorter than the gaseous damping time
and the Larmor precession period, and (2) charge fluctuations have small amplitude
(i.e., $\delta Q \ll Q_{0}$) such that these fluctuations can be approximated
as being continuous and modeled by a Gaussian distribution.
These assumptions are usually not realistic for very small grains for which
the charge fluctuations are substantial. Therefore, the unrealistically high velocities
of very small grains were obtained. This motivates our present study which is intended to
determining the actual velocities of PAHs in typical astrophysical conditions.

In this paper, we aim to quantify the acceleration of very small dust grains
induced by grain charge fluctuations using numerical simulations.
The structure of the paper is as follows. We first discuss major processes
responsible for grain charging and charge fluctuations in \S 2. In \S 3 we present
an algorithm to simulate the charge fluctuations using Monte Carlo method.
In \S 4 we present a numerical simulation method to investigate grain
acceleration, including dynamical equations and Monte-Carlo simulations.
Velocities of very small grains in the ISM from numerical simulations
are presented in \S 5. Discussions and summary are presented in
\S 6 and 7, respectively.

\section{Grain Charging and Charge fluctuations}

Charging processes for a dust grain in the ISM consist of its sticking collisions
with charged particles in plasma (Draine \& Sutin 1985) and photo-emission induced
by UV photons (Draine \& Weingartner 2001). In the former case, the grain acquires
charge by capturing electrons and ions from the
plasma, while, in the latter case, the grain looses charge by emitting photoelectrons.
After a sufficient long time, these processes result in statistical equilibrium
of ionization, and the grain has a mean charge, denoted by \Zmean.
The grain charge fluctuates around the equilibrium value \Zmean.

To characterize the fluctuations of grain charge, we are interested in
the charge distribution function $f_{Z}(Z)$, which describes the probability
of finding the grain with a charge $Ze$.\footnote{This distribution function
also represents the fraction of grains in a charge state because
the grain density in the ISM is so low that each grain does not interfere
with the charging processes of the neighboring grains.} Detailed calculations
for $f_{Z}$ are presented in Draine \& Sutin (1985) and Weingartner \& Draine (2001),
here we describe them briefly. 

Basically, in a steady state approximation, the
distribution function $f_{Z}$ of grain charge is constrained by
the statistical equilibrium
\bea
f_{Z}(Z)\left[J_{pe}(Z)+J_{ion}(Z)\right]=f_{Z}(Z+1)J_{e}(Z+1),\label{eq:fZ}
\ena
which means, the number of positively charged particles the grain
absorbs per second to change the charge state from $Z$ to $Z+1$
must be equal to the number of electrons that the grain absorbs per second
to cascade from $Z+1$ to $Z$. Above $J_{e}$ and $J_{ion}$ are the rate of
sticking collisions with electron and ion, and $J_{pe}$ is the
rate of emission of photoelectron induced by UV photons (see Appendix A).
To find $f_{Z}$, we solve Equation (\ref{eq:fZ}) using an iterative method
as in Draine \& Sutin (1987).

Let us define a characteristic relaxation time of the charge fluctuations,
$\tau_{Z}$, which is equal to the time required for the grain charge to
relax from $Z$ to the equilibrium state (Draine \& Lazarian 1998b):
\bea
\tau_{Z}=\frac{\langle (Z-\langle Z\rangle)^{2}\rangle}{\sum_{Z}f_{Z}J_{tot}(Z)}\equiv\frac{\sigma_{Z}^{2}}{\sum_{Z}f_{Z}J_{tot}(Z)},
\ena
where $J_{tot}(Z)=J_{e}+J_{ion}+J_{pe}$ is the total charging rate at
the charge state $Z$. Here we averaged over all possible charge states $Z$
to find $\tau_{Z}$.

\begin{table}[htb]
\begin{center}
\caption{Idealized Environments For 
 Interstellar Matter}\label{ISM}
\begin{tabular}{llll} \hline\hline\\
\multicolumn{1}{c}{\it Parameters} & \multicolumn{1}{c}{CNM}& 
{WNM} &{WIM}\\[1mm]
\hline\\
$n_{\H}$~(cm$^{-3}$) &30 &0.4 &0.1 \\[1mm]
$T_{\gas}$~(K)& 100 & 6000 &8000 \\[1mm]
$\chi$ &1 &1 &1 \\[1mm]
$x_{\H}$ &0.0012 &0.1 &0.99 \\[1mm]
$x_{\rm M}$ &0.0003 &0.0003 &0.001 \\[1mm]
\hline\hline\\
\end{tabular}
\end{center}
\vspace*{-0.5cm}
Here $n_{\H}$ is the gas density, $T_{\gas}$ is the gas temperature,
$x_{\H}$ is the hydrogen ionization fraction, $x_{\rm M}$ is the metal ionization
fraction, $\chi$ is the ratio of radiation energy density to the
interstellar radiation energy density.
\end{table}

\begin{figure}
\includegraphics[width=0.48\textwidth]{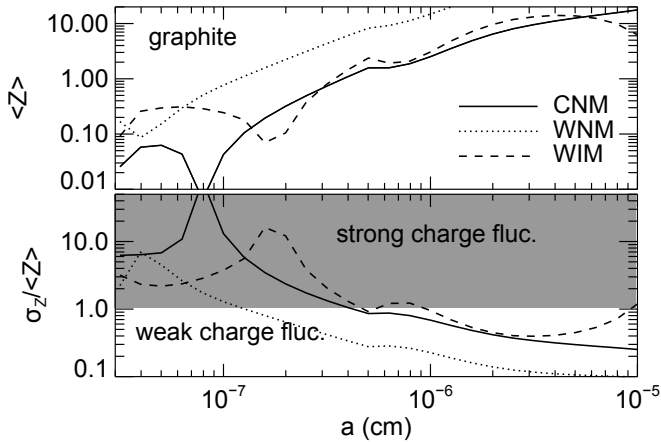}
\caption{Mean grain charge $\langle Z\rangle$ ({\it upper}) and charge dispersion $\sigma_{Z}$ ({\it lower}) of graphite
grains as a function of grain size $a$ for various
ISM phases. Strong charge fluctuations are observed for grains smaller than
$a_{\cf}\sim 10^{-6}$ cm for the CNM and WIM, and $a_{\cf}\sim 10^{-7}$ cm for the WNM.}
\label{fig:sigmaZ}
\end{figure}

Figure \ref{fig:sigmaZ} shows the grain mean charge ({\it the upper plot})
and charge dispersion $\tilde{\sigma}_{Z}=\sigma_{Z}/\langle Z\rangle$
({\it the lower plot}) as a function of the grain size for the cold neutral
medium (CNM), warm neutral medium (WNM) and 
warm ionized medium (WIM). Physical parameters for these phases,
including the gas density $n_{\H}$, gas temperature $T_{\gas}$, radiation
field $\chi$, hydrogen ionization fraction $x_{\H}$ and metal ionization
fraction $x_{\rm M}$ are given in Table 1. The gray area indicates the
region where the charge fluctuations are important with $\tilde{\sigma}_{Z}\ge 1$.
Let $a_{\cf}$ be the grain size corresponding to $\tilde{\sigma}_{Z}=1$.
Figure \ref{fig:sigmaZ} shows that charge fluctuations can be important
for $a<a_{\cf}$, where $a_{\cf}\sim 10^{-6}~\cm$ for the CNM and WIM, and
$a_{\cf}\sim 10^{-7}~\cm$ for the WNM.

\begin{figure}
\includegraphics[width=0.5\textwidth]{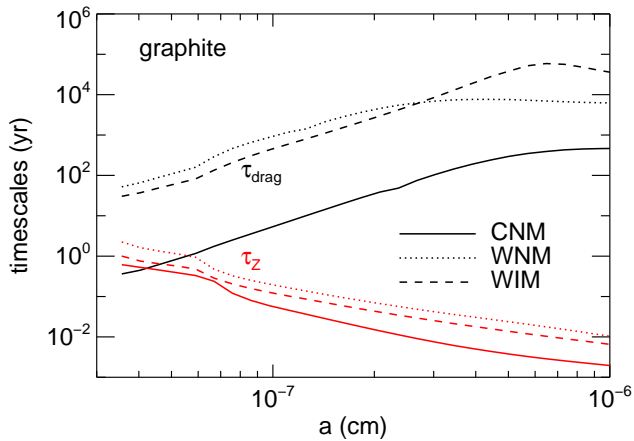}
\caption{The gas drag timescale $\tau_{\drag}$ and the relaxation time of
charge fluctuations $\tau_{Z}$ as
functions of the grain size $a$ for graphite grains in the various
phases of the ISM. Charge fluctuations are slow with $\tau_{Z}\leq \tau_{\drag}$
for $a<5\times 10^{-8}~\cm$ in the CNM.}
\label{timescale}
\end{figure}

We calculate the relaxation time of charge fluctuations $\tau_{Z}$ and the gas
damping time $\tau_{\drag}$ (Eq. \ref{eq:tau_drag}) for both
graphite and silicate grains
in various phases of the ISM with physical parameters listed in Table 1. Obtained results
are shown in Figure \ref{timescale}. In the WNM and WIM, the charge fluctuations are fast
(i.e.,  $\tau_{Z}\ll \tau_{\drag}$)
for the entire size distribution. In the CNM, the charge fluctuations are slow, i.e. $\tau_{Z}$
comparable to $\tau_{\drag}$ for $a<10^{-7}$ cm. The slow charge fluctuations arise from
the fact that the ionization fraction is rather low in the CNM, which makes the charging much
less frequent than the gas-grain collisions.

\section{Monte Carlo Simulations of charge fluctuations}

For very small grains (e.g., PAHs) under interest in the ISM, the charging is infrequent,
so that the charge fluctuations may not be adequately characterized by the steady state
distribution $f_{Z}$. In addition, what we are interested in the present paper is the
time-dependent
grain charge, instead of the steady distribution. Thus, in the following, we first describe an algorithm
to simulate the grain charge fluctuations using the Monte Carlo
method and find the time-dependent charge. Then, we compare the resulting charge distribution function
with the steady state charge distribution obtained using the statistical
equilibrium approximation.

\subsection{Monte Carlo Simulations}
Monte Carlo simulations of the charge fluctuations for dust grains in
plasma have been studied in Cui \& Goree (1994), but they
considered only the charge fluctuations arising from collisions with electrons
and ions in the plasma. Here, we consider a general case of charging arising
from both sticking collisions and photoemission induced by UV photons.

\begin{figure}
\includegraphics[width=0.5\textwidth]{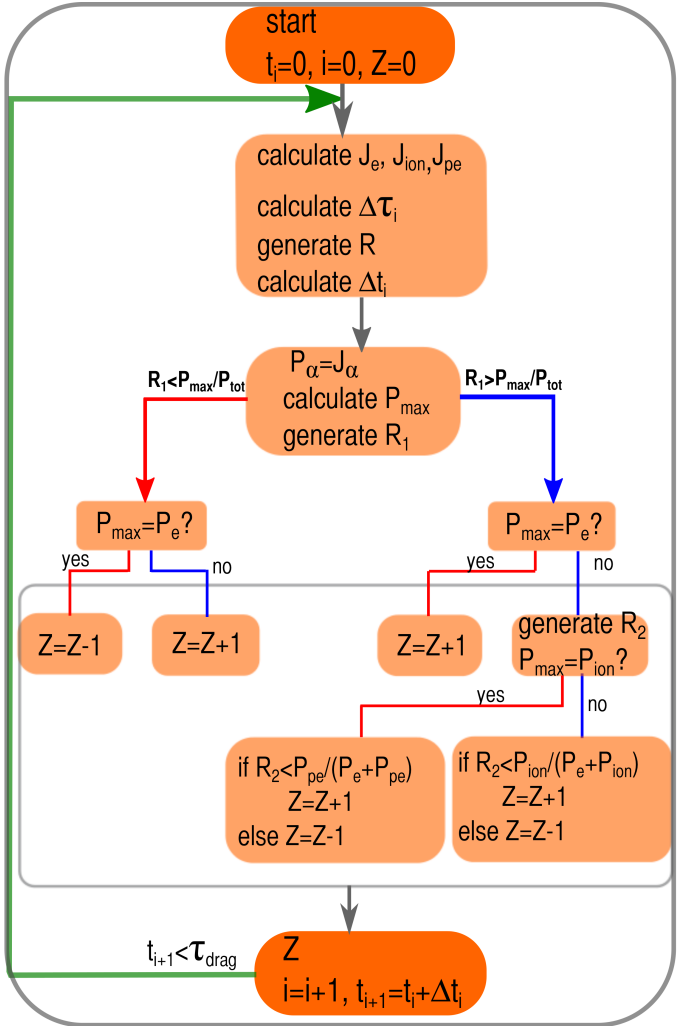}
\caption{Diagram for MC simulations of charge fluctuations. Initially, grain is neutral.
The charging rates $J_{e},~J_{ion}$ and $J_{pe}$ are used to calculate the
mean time between two charging events $\Delta \tau$ and the waiting time to the
next charging event $\Delta t$. The simulations are iterated until the elapsed
time $t_{i+1}$ exceeds the drag time $t_{\drag}$.}
\label{MC_alg}
\end{figure}

The underlying idea of MC simulations is, that, we identify the moment of
the charging events that occur randomly,\footnote{A charging event can be
the absorption of an electron (ion) or emission of a photoelectron.} and assume
that the grain charge changes instantly at the moment of charging. We keep
track of the grain charge and the time interval between two successive
charging events. The algorithm for simulating the charge fluctuations
is presented in Figure \ref{MC_alg}. At the initial step $i=0$ and time $t=0$,
the grain is assumed to be neutral with charge $Ze=0$.
We calculate the rate of sticking collisions with electrons and ions,
$J_{e}(Z),~ J_{ion}(Z)$, and the rate of emission of photoelectrons, $J_{pe}(Z)$. The total charging rate reads $J_{\rm tot}(Z)=J_{e}+J_{ion}+J_{pe}$.
The waiting time to the next $i+1$ charging event is a random variable
drawn from a Gamma distribution function with the mean value
${\Delta \tau_{i}}=1/J_{\rm tot}(Z)$.

In fact, the probability to have a charging event occurring in $[t,t+dt]$ is
\bea dP=
\Delta\tau_{i}^{-1}\exp\left(-t/\Delta\tau_{i}\right)dt.
\ena
The probability to find the next charging event $i+1$ after a time interval
$\Delta t_{i}$ becomes
\bea
P=1-\exp\left(-\frac{\Delta t}{\Delta \tau_{i}})\right).\label{eq:Prob}
\ena
To find $\Delta t_{i}$, we generate a random number $R$ in the range [0,1],
and assign $P\equiv R$. From Equation (\ref{eq:Prob}) we obtain
\bea
\Delta t_{i}=\frac{\ln (1-R)}{\Delta \tau_{i}}.
\ena

When the waiting time to the next charging event $\Delta t_{i}$ is known,
we have to determine what particle among $e,~ion$, and photoelectron (pe)
the grain captures or emits at the next charging event. We note that the
probability of capturing or emitting $\omega$ particle is $P_{\omega}=J_{\omega}$
with $\omega=e,~ion$ and $pe$. The probability that the next
charging event corresponds to the absorption/emission of the $x$ particle
among $e,~ion$ and $pe$ is given by $P_{x}/P_{\tot}$ where
$P_{\tot}=P_{e}+P_{ion}+P_{pe}$. The next charging event corresponds to
the absorption/emission of the $x$ particle when $P_{x}$ is maximum.

We begin with generating a random number $R_{1}$ in the range $[0,1]$
from a uniform distribution. If $R_{1}<P_{\max}/P_{\tot}$, then we know that
a charging event corresponding to $P_{\max}$ just occurred. Now, if
$P_{\max}=P_{e}$, then one electron is captured by the grain in this charging
event, and the grain charge $Z$ is advanced by $-1$. If $P_{\max}\ne P_{e}$,
then the grain captures either an ion or emits a photoelectron in this
charging event. For this case, $Z$ is advanced by $+1$.

If $R_{1}>P_{\max}/P_{\tot}$, then we know that the next charging event
with $P_{\max}$ has not occurred, and the charging event occurred with
one of two remaining particles. To determine what particle the grain
actually absorbs/emits, we continue by generating a random number $R_{2}$,
and compare it with $P$ of these particles. The grain charge $Z$ is updated,
and the time is advanced as $t_{i+1}=t_{i}+\Delta t_{i}$. We tabulate the
charging time $t_{i+1}$ and grain charge $Z(t)$.
This process is iterated until the elapsed time $t_{i+1}$ equal to a few gaseous
damping times $\tau_{\drag}$.

\begin{figure}
\includegraphics[width=0.5\textwidth]{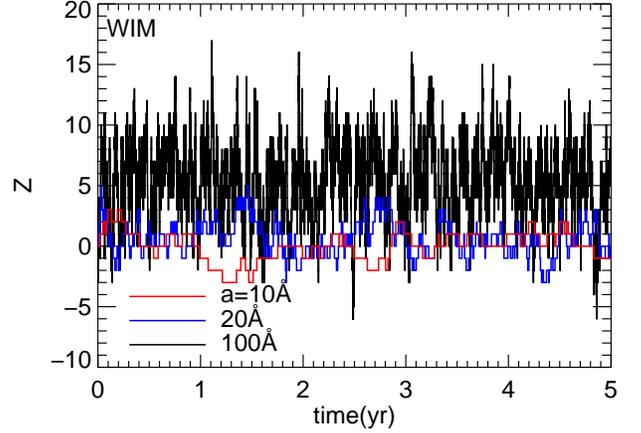}
\caption{Grain charge as a function of time obtained from MC simulations for three
values of grain size $a=10\Angstrom$ (red line), $~20\Angstrom$ (blue line),
and $100\Angstrom$ (black line) in the WIM. Charge fluctuations are rather
slow for the smallest grains (red and blue lines), but fast for larger grain (black line).}
\label{f3}
\end{figure}

Figure \ref{f3} shows the time-dependent grain charge obtained using MC simulations
for a $10,~20$ and $100\Angstrom$ grain in the WIM. The discrete nature can
be easily seen for very small grain $a=10\Angstrom$ in which $Z$ varies slowly.

\subsection{Charge distribution function}

\begin{figure}
\includegraphics[width=0.5\textwidth]{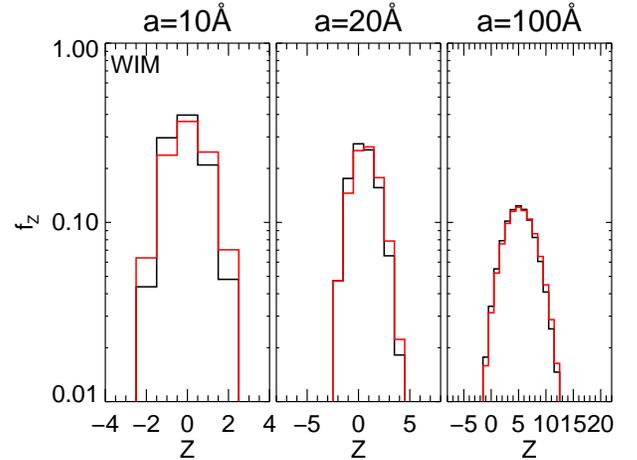}
\caption{Charge distribution function $f_{Z}$ obtained from MC simulations
(red line) compared to $f_{Z}$ from the statistical equilibrium
approximation (black line) for three values of grain size in the WIM. $f_{Z}$
from two approaches is almost similar for the $100\Angstrom$ grain
when the charge fluctuations are fast, but there is some difference for
two smaller sizes for which the charge fluctuations are slow.}
\label{fZ_comp}
\end{figure}

To characterize the fluctuations of grain charge from MC simulations,
we find the steady state charge distribution function $f_Z$ using the
tabulated data $Z(t)$. Figure \ref{fZ_comp} compares $f_{Z}$ obtained
from MC simulations with $f_{Z}$ obtained using the statistical equilibrium
approximation (see Eq. \ref{eq:fZ}) for different grain sizes in the ISM.
For $a=100\Angstrom$ grains, the results from two different approaches
are similar, while there exists some difference in $f_{Z}$ for smaller grains
(e.g., $a=10$ and $20\Angstrom$).
Such a difference stems from the fact that the charge fluctuations occur
at much lower rate for very small grains, so that the statistical
equilibrium approximation is not adequate.

\section{Numerical simulations of Grain acceleration}
\subsection{Basic Equations}
Let consider an ensemble of grains in a plasma with density $n_{H}$,
ionization fraction $x_{\H}$, and temperature $T_{\gas}$. The velocity of
a grain $i$ of mass $m_{i}$ changes due to the gas drag force as well as
the Coulomb force induced by charged grains in its vicinity. The conventional
equation of motion reads
\bea
\frac{d\bv_{i}}{dt}=-\frac{\bv_{i}}{\tau_{\drag}}+R_{i}+\frac{\bFr_{i}}{m_{i}},\label{eq:dvdt}
\ena
where $\tau_{\drag}$ is the damping time due to gas drag (see Eq.\ref{eq:tau_drag}),
$R_{i}$ is a random force due to Brownian motions, which is described by
\bea
\langle R_{i}^{2}\rangle=G_{i},
\ena
with $G_{i}=2\tau_{\gas}^{-1}k_{\B}T_{\gas}$ being the translational
excitation coefficient due to dust-gas collisions. Here $\bFr_{i}$ is the
Coulomb force acting on $i$-grain induced by all surrounding grains
$j\ne i$. Assuming the Yukawa potential with the screened length
$\lambda$, $\bFr_{i}$ is given by
\bea
\bFr_{i}(\br,t)&=&\sum_{j\ne i}\nabla \frac{Q_{i}(t)Q_{j}(t){\exp}\left(-r_{ij}/\lambda\right)}
{\br_{ij}}\nonumber\\
&=&\sum_{j\ne i}Q_{i}(t)Q_{j}(t)\frac{\exp(-r_{ij}/\lambda)}{r_{ij}^2}
\left(\frac{1}{\lambda}+\frac{1}{r_{ij}}\right)\br_{ij},~~~\label{eq:force}
\ena
where $Q_{i}(t)$ and $Q_{j}(t)$ are the charge of $i$ and $j$ grains at the
time $t$, and $r_{ij}$ is the separation between them.

The position of the $i$ grain is determined by the equation
\bea
\bv_{i}=\frac{d\br_{i}}{dt}.
\ena
Solving Equation (\ref{eq:dvdt}) is rather challenging, because it is involved
with the stochastic term $R_{i}$ and a time-dependent force $\bFr$,
which depends on the instant grain charge $Q(t)$.

\subsection{Numerical Integration}

To solve Equation (\ref{eq:dvdt}) combined with Equation (\ref{eq:force})
for grain velocities, we first need to know the fluctuating charge $Q(t)$ at
the time $t$. Detailed treatment of $Q(t)$ using data from MC simulations
will be addressed in the next section. Here, we present a general numerical
approach to solve Equation (\ref{eq:dvdt}) for an arbitrary time-dependent force
$\bFr(\br,t)$.

Let us denote $h_{n}$ be the time interval at the step $n$ in which the
charge of all grains are constant. The velocity of a grain at the step $n+1$
is given by
\bea
\tilde{\br}_{n+1}&=&\br_{n}+\bv_{n} h_{n},\label{eq:rnp}\\
\bv_{n+1}&=&\bv_{n}+\left[-\frac{\bv_{n}}{\tau_{\drag}}+R_{i} +
\frac{\bFr_{i}+\bFr_{i}'}{2m_{i}}
 \right]h_{n},\label{eq:vn}\\
\br_{n+1}&=&\br_{n}+\frac{1}{2}\left(\bv_{n}+\bv_{n+1}\right)h_{n},~\label{eq:rn}
\ena
where $\bFr_{i}=\bFr_{i}(\br_{n},t_{n})$ and $\bFr_{i}'=\bFr_{i}(\tilde{\br}_{n+1},t_{n})$. 

In simulations we adopt dimensionless units
\bea
t'=\frac{t}{\tau_{\drag}},~~ {\bv}'=\frac{\bv}{v_{T}},\label{vnorm}
\ena
where $v_{T}=\left(k_{\B}T_{\gas}/m_{i}\right)^{1/2}$ is the grain thermal velocity. The position
is normalized over the screened length scale $\lambda$:
\bea
\br'=\frac{\br}{\lambda}.\label{rnorm}
\ena

Using the dimensionless units, Equations (\ref{eq:rnp})-(\ref{eq:rn})
can be rewritten as
\bea
\br'_{n+1}&=&\br'_{n}+\bv'_{n} h'_{n},\label{rnp}\\
\bv'_{n+1}&=&\bv'_{n}+\left[-\bv'_{n}+R'_{i} +
\frac{\bFr_{i}+\bFr'_{i}}{2m_{i}}\frac{\tau_{\drag}}{v_{T}\lambda^{2}}
\right]h'_{n},\label{vn}\\
\br'_{n+1}&=&\br'_{n}+\frac{1}{2}\left(\bv'_{n}+\bv'_{n+1}\right)h'_{n}. \label{rn}
\ena

When the velocities of grains is known, the kinetic temperature is evaluated by
the ensemble average:
\bea
\frac{3k_{\B}T_{d}}{2}=\frac{1}{N_{p}}\sum_{1}^{N_{p}} \frac{m_{d}v^{2}}{2}.
\ena

In the above equations, we note that the grain charge $Q$ changes in time.
Therefore, to obtain grain velocity at each timestep, we need to treat $Q(t)$ properly.
Below, we present an algorithm to find $Q(t)$ at each timestep using data
from MC simulations.

\subsection{Algorithm}
\subsubsection{Tiny grains $a<10\Angstrom$}

Very small grains with $a\leq 10\Angstrom$ have infrequent charging
(see Fig. 4), thus most of the time they spend in the charge states
$Z=0,\pm 1$, and $\pm2$.

Since the grain charge $Q$ is a function of time, the time step $h_{n}$
in Equation (\ref{rnp})-(\ref{rn}) must be chosen such that the charge of
grains in the simulation box remains constant during this timestep. The easy
way is to choose the timestep coincident with the charging time
of all grains.

We consider $N_{p}$ grains in the simulation box. For each grain, we
generate $N_{Z}$ random charging events, and tabulate the moment of
charging $t_{Z}$ and the time interval between two charging events $\Delta t_{Z}$
using MC simulations (here $t_{Z}\equiv t_{i}$ in Section 3). As a result,
we have total $N_{\chrg}=N_{p}\times N_{Z}$ charging events in the
entire simulation box. When the charging moment $t_{Z}$ is known,
we can sort it in ascending order, and obtain $t_{\chrg}^{n}$ with
$n=1-N_{\chrg}$ for the sequence of $N_{\chrg}$ charging events in the simulations.
The timestep is then chosen as the time interval between two successive
charging events in the sequence, i.e., $h_{n}=t_{\chrg}^{n}-t_{\chrg}^{n-1}$.
When sorting $t_{\chrg}$, we also mark the grain that experiences the charging
event. Therefore, at each time step, only this {\it marked} grain has charge
that changes instantly by $\pm 1$,  while the charge of other grains remains
the same. When the integration time $t_{n}$ is equal to the charging
time $t_{Z,i}$, then the charge of $i$ grain changes.

If the timestep $h_{n}$ results in a large increment of $v_{n}$, then we need to
adjust $h_{n}$ so that the increment of grain velocity is not larger than an
$\epsilon$ fraction of $v_{n}$. We choose $\epsilon=0.1$ for our
simulations. When the time interval between charging events is very large,
we need to limit $h_{n}^{\max}=10^{-5}$ to keep the numerical integration stable.

\subsubsection{Very small grains $a\ge10\Angstrom$}

For grains larger than $10\Angstrom$, the charging occurs more frequently,
so that the time interval between two successive charging events $dt_{\chrg}$
may be rather small (about $10^{-9}\tau_{\drag}$). As a result, we need to
integrate Equation (\ref{eq:dvdt}) over a huge number of timesteps
to achieve the terminal velocity after $\tau_{\drag}$, which is rather impractical.
To overcome this obstacle, we choose a constant timestep $h_{n}$, and
employ the MC method to generate the instant charge at each timestep.

The basic idea is that, the instant charge $Z$ is randomly distributed, and can be drawn
from the steady state distribution $f_{Z}$. We first calculate the accumulative probability
of finding the grain with charge $\le Ze$ as
\bea
P(Z)=\sum_{Z_{i}=Z_{\min}}^{Z} f_{Z}(Z_{i}),\label{pz}
\ena
where $Z_{\min}$ is the minimum grain charge. The grain charge $Z$
at each timestep is random, and obtained
by solving equation $P(Z)=R$ where
$R$ is a random variable drawn from a uniform distribution.

\subsection{Boundary and initial conditions}

We adopt the periodic boundary conditions for our simulations. The periodic
boundary assures that when a dust grain leaves the box at one boundary,
another grain is added to the box at its opposite boundary. The periodic
boundary is useful for simulating large scale systems consisting of a huge
number of particles.

We start simulations with $N_{p}$ particles randomly distributed in a square
box of $L^{2}$. Their initial velocities are drawn from a Gaussian distribution
with a kinetic temperature $T_0$.

\subsection{Code benchmarking: Brownian motion}
We test the code by simulating the Brownian motions of dust grains in a
plasma of temperature $T_{\gas}$. We consider an ensemble of $N$ neutral
grains of equal size in the conditions of the CNM and WIM. Simulations
are started with an initial temperature $T_{0}<T_{\gas}$. We found that
the grain kinetic energy is driven to thermal equilibrium with the plasma
after about the gaseous damping time $\tau_{\drag}$.

\section{Grain acceleration due to charge fluctuations in the ISM}

\subsection{Model parameters}

Below we are going to find the actual velocities of very small grains
induced by charge fluctuations in various phases of the ISM, which have
physical parameters listed in Table 1. We consider the translational damping
due to grain-neutral and grain-ion collisions only (see Appendix B), and
disregard subdominant damping processes (e.g., dipole-dipole,
grain-dipole interactions).
We also disregard the sticking grain-grain collisions due to the fact that
the mean distance between grains in the ISM is large, which is comparable
to the Debye length (see Ivlev et al. 2010).

\subsection{Grain size distribution}

First, let us assume for the sake of simplicity that all dust grains have
the same equivalent size $a_{\rm eq}$ with density $n_{\d}(a_{\rm eq})$.
Here, the density $n_{\d}(a_{\rm eq})$ is constrained by
the condition that the mass density of $a_{\rm eq}$ grains is equal
to the total mass density of dust integrated over the entire size distribution from $a_{\min}$ to $a_{\max}$:
\bea
M_{\d}\equiv\frac{4\pi}{3}a_{\rm eq}^{3}n_{\d}(a_{\rm eq})=\int_{a_{\min}}^{a_{\max}} \frac{4\pi}{3}a^{3}\frac{dn}{da}da,
\ena
where $a_{\min}=3.56\Angstrom$ and $a_{\max}=2.5\times 10^{-5}$ cm are assumed.
Assuming the MRN distribution $dn/da=n_{\H}A_{\rm MRN}a^{-3.5}$ with $A_{\rm MRN}=10^{-25.16}\cm^{-2.5}$ for diffuse clouds (see Mathis et al. 1977), we obtain
\bea
n_{\d}(a_{\rm eq})\approx10^{-6}n_{\H}\left(\frac{a_{\rm eq}}{10^{-7}\cm}\right)^{-3}.\label{nd_eq}
\ena

We perform MC simulations for different equivalent sizes $a_{\rm eq}$
from $a_{\min}$ to $a_{\max}$. For each value of $a_{\rm eq}$, the density
is inferred from Equation (\ref{nd_eq}).

Next, we account for the electrostatic interactions between grains of different sizes.
We divide the grain size distribution from $a_{\min}$ to $a_{\max}$ into $N_{\rm bin}$
with a constant log binsize, $\Delta \ln a=\ln(a_{\max}/a_{\min})/(N_{bin})$.
The grain size in the $i$ bin is given by
\bea
\ln a_{i}=\ln a_{\min}+i\Delta \ln a,
\ena
and the fraction of grains in $i$ bin to the total dust density $n_{d}$ reads
\bea
\frac{n_{i}}{n_{d}}\equiv \frac{(dn_{i}/da)\Delta a_{i}}{n_{d}}=\frac{a_{i}^{-3.5}a_{i}}{\sum_{k=0}^{k=N_{\rm bin}-1}a_{k}^{-3.5}a_{k}},\label{eq:na}
\ena
where we have adopted the MRN distribution for simplicity.

Assuming the total number of particles in the simulation box is $N_{p}$,
the box size $L$ is constrained by
\bea
\sum_{i=0}^{N_{\rm bin}-1} n_{i} L^{2}=n_{d}L ^{2}=N_{p}.
\ena

We run simulations for $N_{p}$ grains of various size $a$, and find the
average velocity for grains having the same size.  For each grain size $a$,
the dominant contribution to the acceleration arises from its interaction
with other grains of size $a'>a$. Therefore, the Coulomb force in
Equation (\ref{eq:force}) corresponds to the summation over all grains
larger than the grain of mass $m_{i}$ under interest.

\subsection{Grain Velocities}
We present the results of MC simulations for first the simplest case in
which all grains have the same equivalent size. For each grain size,
we follow the time evolution of grain velocity $v$ and calculate
the averaged velocity $v$ at each time step. We identify the time interval in
which $v$ varies slowly and then saturates. The terminal
velocity is obtained by averaging $v$ over the time interval
ranging from $\tau_{\drag}$ to the total integration time $T$.
Here, we adopt the total integration time $T=2\tau_{\drag}$.

\begin{figure}
\includegraphics[width=0.5\textwidth]{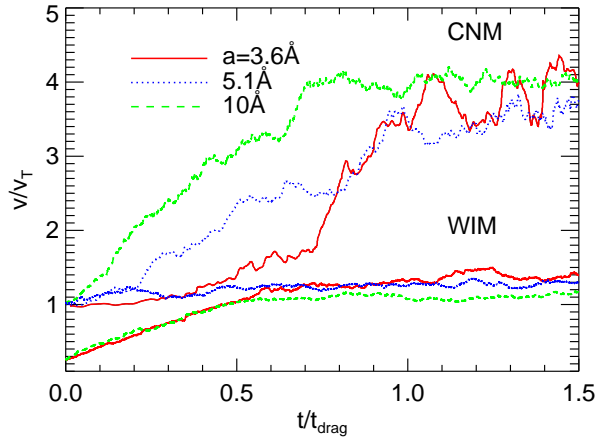}
\caption{Evolution of grain velocity normalized to its thermal velocity as
a function of time from MC simulations for three grain sizes in the CNM
and WIM. Graphite grains are considered. Charge fluctuations
accelerate very small grains in the CNM to suprathermal velocities after about
the gas drag timescale $\tau_{\drag}$. }
\label{Et_cnm}
\end{figure}

Figure \ref{Et_cnm} presents the temporal evolution of grain velocity normalized
over its thermal velocity in the CNM and WIM for three grain sizes.
As expected, the grain velocity increases exponentially as a function of
time from the initial value, then slows down gradually and saturates after
about a gas damping time. The terminal velocity is obviously larger for
smaller grains. In addition, we can see that the acceleration by charge fluctuations
is more efficient in the CNM than the WIM, which
can accelerate grains up to about four times its thermal velocity.
The reason for that is as follows. First, the CNM has stronger charge fluctuations,
which corresponds to larger charge dispersion
$\sigma_{Z}$/\Zmean~(see Fig. \ref{fig:sigmaZ}), resulting in stronger
fluctuating electric fields. Second, since the CNM has lower degree of
ionization, which corresponds to larger Debye length, the shielding effect of
plasma on electrostatic interactions between charged grains is much less
than in the WIM.

\begin{figure}
\includegraphics[width=0.48\textwidth]{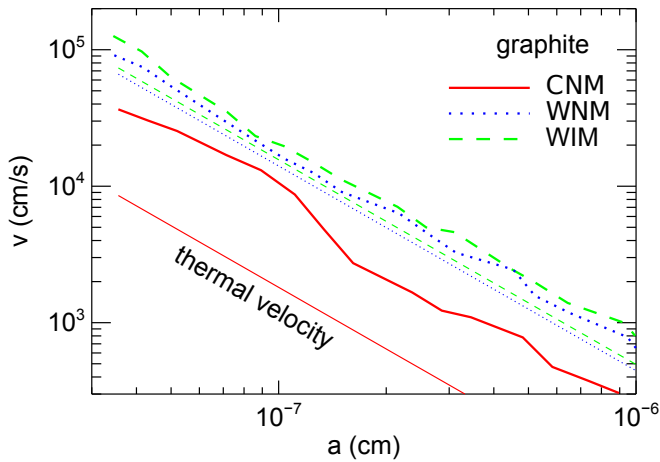}
\caption{Terminal grain velocities as a function of grain size for the CNM,
WNM and WIM due to charge fluctuations are shown by solid, dotted and
dashed lines, respectively. Thin lines show the thermal velocities of
grains in the corresponding ISM phases.}
\label{vd_cnm}
\end{figure}

To see the dependence of the terminal velocity with grain
size $a$ clearly, in Figure \ref{vd_cnm}, we show results for three phases
of the ISM. We truncate at the size $a=10^{-6}$ cm because the charge
fluctuations are negligible for large grains. As shown
is the rapid increase of velocity with decreasing grain size.

\begin{figure}
\includegraphics[width=0.48\textwidth]{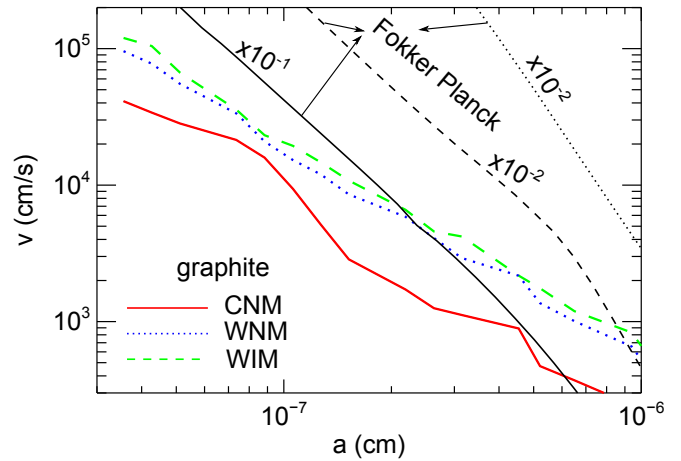}
\caption{Similar to Fig \ref{vd_cnm}, but the effects of grain size distribution
are taken into account. Grain velocities predicted by the Fokker-Planck
equation are much larger than the actual values computed by MC simulations.}
\label{vdsize_cnm}
\end{figure}

To study the effect of grain size distribution on grain acceleration,
we perform simulations for 16 grain sizebins from $a=3.56\Angstrom$ to
$100\Angstrom$ and take into account the electrostatic interactions of
grains of different sizes. Obtained velocities for the CNM, WNM, and WIM
are shown in Figure \ref{vdsize_cnm}.
As shown, the velocities of the smallest grains increase while the velocities
of larger grains decrease slightly, compared to the velocities obtained for
the grains of equal size. That is understandable because, when the grain size
distribution is taken into account, the smallest dust grains (PAHs) have
chance to interact with larger grains. Largest grains, on the contrary, interact
mostly with smaller grains so that their velocities decrease.

\subsection{Comparison with Fokker-Planck Equation Approach}

The Fokker-Planck (FP) equation was used in Ivlev et al. (2010) to study
the grain acceleration by charge fluctuations.
The key assumptions in the approach based on the FP equation
are as follows: (1) grain charge fluctuations are fast and (2) the charge
dispersion is small compared to the equilibrium charge. These assumptions
are, in general, valid for large grains, but they fail for very small grains
(see Fig. \ref{fig:sigmaZ}). To see the consequence of such unrealistic
assumptions on predicting PAH velocities,
we conservatively apply the FP approach for very small grains and compute
grain velocities using Equation (\ref{vd}) in which
the actual values of $\sigma_{Z}$
and \Zmean~from Figure \ref{fig:sigmaZ} are adopted.

Figure \ref{vdsize_cnm} compares grain velocities obtained from MC simulations
with those predicted by the FP approach for the ISM. We can see that
the FP approach predicts much larger grain velocities than the actual
values from MC simulations. The difference in grain velocities from the
two approaches decreases when $a$ increases and becomes
negligible for sufficient large grains.

The reason for that the FP approach predicts unrealistically
high velocities for PAHs is the following. When adopting the FP approach
to estimate grain velocities, we are implicitly
assuming an unrealistic situation in which PAHs have fast fluctuating
charges with large amplitude.
It corresponds to the situation in which each PAH is subject to unusually strong
and fast fluctuating electric fields. As a result,
PAHs gain a substantial amount of energy after a short interval of time
and are accelerated to extremely high velocities. However, for very small grains (e.g., PAHs),
we showed in Figures 1 and 2 that the grain charge indeed has large dispersion, but
it fluctuates rather slowly, which makes the acceleration much less efficient
than the prediction by the FP equation.

For highly ionized media, the fast charge fluctuations can be applied,
but we found that the results are still much lower than the FP theoretical
prediction. Such a difference can arise mainly from the fact that the plasma screening
effect is disregarded in the estimate of Ivlev et al. (2010), while the highly ionized
media have short Debye lengths, which reduce the electrostatic dust-dust
interactions.

\section{Discussion}

\subsection{Grain Acceleration from Various Mechanisms}
Grain acceleration by hydrodrag (see Draine 1985) was discussed
as the dominant mechanism driving grain motions in the ISM as well
as in protoplanetary disks. Recently, thanks to significant progresses in
understanding of MHD turbulence, Lazarian \& Yan (2003),
YLD04, and Yan (2009) identified the gyro-resonant interactions of grains
with fast modes in MHD turbulence as the dominant mechanisms to
accelerate large grains ($a>10^{-5}$ cm) to super-Alfv'{e}nic velocities.
For very small grains (e.g., PAHs and nanoparticles), the gyro-resonance
acceleration is inefficient for most phases of the ISM, because PAHs and
nanoparticles with smaller gyro radii only resonantly interact with small
scale and weak fluctuations.
Hoang et al. (2011b) revisited the treatment of gyroresonance acceleration
by MHD turbulence by accounting for the fluctuations of the grain guiding
center with respect to the mean magnetic field. They improved estimates
of the previous authors, and proposed a new way of acceleration through
transit time damping (TTD) of fast modes, which is efficient for
super-Alfv\'{e}nic grains (see solid lines in Fig. \ref{fig_TTD}).

\subsection{Acceleration induced by Charge Fluctuations}
In the present paper, we have numerically investigated the
grain acceleration mechanism induced by charge fluctuations
for very small grains. The combination of MC simulations for
charge fluctuations with direct simulations of electrostatic interactions
between grains with fluctuating charge allows us to follow the
evolution of grain velocities in time and estimate their terminal
velocities. We showed that the acceleration induced by charge
fluctuations is indeed an important mechanism to accelerate PAHs.
Resulting grain velocities can be several times higher than their
thermal velocities, which is obviously important.

\begin{figure} [h]
\includegraphics[width=0.49\textwidth]{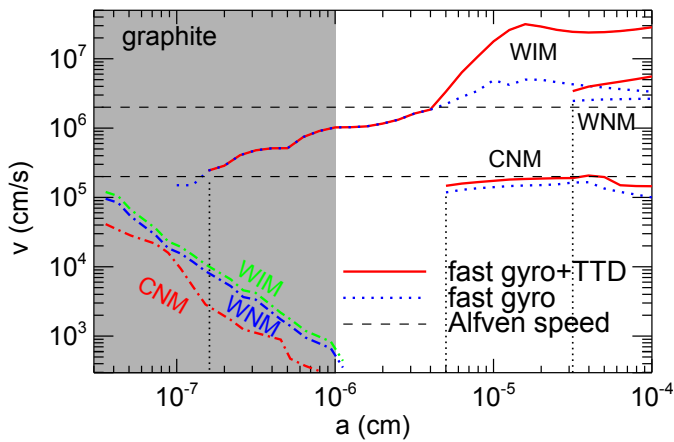}
\caption{Grain velocity as a function of grain size $a$ due to charge fluctuations
(shaded area) and due to resonance 
acceleration by MHD turbulence (Hoang et al. 2011b) for the CNM, WNM,
and WIM. MHD turbulence is the efficient mechanism for large grains, and
grain charge fluctuations work efficiently at the small end of grain size. Dotted
and solid lines correspond to the acceleration from gyroresonance and gyroresonance
plus transit time damping.}
\label{fig_TTD}
\end{figure}

We also found that grains velocities obtained from MC simulations are between
two and three orders of magnitude lower than
the prediction by the FP equation. Such overestimate of the FP approach
arises because the charge fluctuations of very small grains
in the ISM are slow, in contrast to the assumption of fast charge fluctuations
assumed in the FP treatment, which makes electric fields
fluctuating rather slowly. In addition, the FP treatment disregarded the
screening effect of plasma. In the ISM, the mean separation between two
grains is comparable to the Debye length.
Thus, the Coulomb interaction is suppressed considerably, and the grain
acceleration is decreased accordingly.

Figure \ref{fig_TTD}
compares grain velocities induced by charge fluctuations (shaded area) with
those due the resonance interactions in MHD turbulence. As shown, MHD
turbulence is efficient for large grains, while charge fluctuations are important for
very small grains.

\subsection{Grain coagulation and shattering}

The grain coagulation is a process in which small dust particles collide
with each other to form aggregates of submicron size (Spitzer 1978).
The grain coagulation can occur in dense clouds (e.g., Ossenkopf 1993;
Ormel et al. 2009) as well as in protoplanetary disks (see van Boekel et al. 2003).

Grain coagulation and shattering result from grain-grain collisions,
which depends on grain relative velocity, denoted by $v_{dd}$.
The threshold velocity for the grain shattering is a function of the grain size:
\bea
v_{\rm shat}=2.7\left(\frac{a}{10^{-7}~\cm}\right)^{-5/6} {\rm km}~\s^{-1},\label{v_cri}
\ena
(Chokshi et al. 1993). If $v_{dd}< v_{\rm shat}$, the grains collide
and stick together. When $v_{dd} > v_{\rm shat}$, the collisions
with high velocity produce shock waves inside the grain, and shatter them
in smaller fragments. For $v_{dd}\rightarrow 20$ km$\s^{-1}$, the evaporation of
dust grains occurs and grains are destroyed.

We note that due to resonant acceleration by MHD turbulence,
large grains $a>10^{-5}$ cm move frequently with velocity
larger than $v_{\rm shat}$ (see Fig. \ref{fig_TTD}). Thus, the shattering
of large grains can be efficient to replenish very small grains into the ISM.
The effect of grain acceleration by MHD turbulence on grain coagulation
and shattering was studied in Hirashita \& Yan (2009). In some phases, they
found that grain acceleration can modify grain size distribution.
At the same time, very small grains move faster than thermal speed
due to the acceleration discussed in the present paper and enhance
grain coagulation. Therefore, future models of grain evolution
should take into account the acceleration of very small grains.

\subsection{Effects of PAH acceleration in protoplanetary disks}

Grain coagulation is the first step to planetesimal formation in
protoplanetary disks (PPDs). It is also believed that the coagulation
from very small grains to micron size occurs rapidly (see e.g., van Boekel et al. 2003),
so that PAHs and nanoparticles are rapidly depleted in PPDs.
However, recent observations by Infrared Space Observatory (ISO) and {\it Spitzer}
satellite reveal PAH emission features, which indicates the existence of PAHs in
young intermediate mass (Herbig Ae/Be) stars (Acke \& van den Ancker 2004)
and in the surface layers exposed to UV radiation in young low mass (T-Tauri) stars
(Geers et al. 2006; Oliveira et al. 2010; Bern\'{e} et al. 2009).

PPDs have highly stratified structure, that means,
there exists some gradient of grain size from the midplane to the surface.
When PAH emission is seen, it is probably only from PAHs in the surface layers
that "see" direct stellar radiation. The abundance of PAHs and other
nanoparticles below the surface is rather uncertain.

{\it (a) Effects of acceleration on dust coagulation and planet formation}

Current models of grain growth in PPDs disregarded grain charges
and their fluctuations. Recently, Okuzumi (2009) took into account the effects of
grain charge and showed that Coulomb potential barrier can suppress
grain coagulation. Okuzumi (2009) appealed to turbulence as an energy source
to overcome the potential barrier. Below, we show that charge fluctuations can
induce very small grains to overcome the Coulomb barrier assuming that PPDs
have a sufficiently high degree of ionization for which the charge fluctuations
are important.

Assuming that the charge fluctuations are Gaussian, then, the mean charge
and charge dispersion are respectively given by
\bea
\langle Z\rangle=z\frac{ak_{\B}T_{\gas}}{e^{2}}\simeq-0.1\left(\frac{a}{10^{-6}\cm}\right)
\left(\frac{T_{\gas}}{100~\K}\right),~~~\label{zmean}
\ena
where $z=2.5$ is adopted (see Morfill \& Ivlev 2009).

The dust mass density in PPDs is related to the gas density as
\bea
\rho_{d}=f_{dg}\rho_{g},\label{rhod}
\ena
where $f_{dg}=0.014$ for the PPD (see Tanaka et al. 2005).
The dust density of the grain size $a$ is then
\bea
n_{d}=10^{1}\left(\frac{a}{10^{-6}~\cm}\right)^{-3}
\left(\frac{n_{\gas}}{10^{10}~\cm^{-3}}\right).\label{nd_ppd}
\ena
\\
Using above equations for Equation (\ref{Tterm}), we obtain
\bea
\frac{T_{d}^{\infty}}{T_{\gas}}=9.31\times 10^{17}\epsilon^{2} \left(\frac{T_{\gas}}{100~\K}\right)^{2}
\left(\frac{a}{10^{-6}~\cm}\right)^{-4}\nonumber\\
\times x_{i} \left(\frac{0.1~\cm^{-3}}{n_{i}}\right)^{2}\left(\frac{n_{\gas}}{10^{10}~\cm^{-3}}\right),~~~~~\label{vd_ppd}
\ena
where $\epsilon<1$ is introduced to account for the overestimate of
grain velocities obtained using the FP approach, $n_{i}$ is the ion density, $x_{i}=n_{i}/n_{\gas}$
is the ionization fraction of gas, and PPDs are
assumed to be isothermal plasma with $T_{\gas}=T_{i}$.

Now, let us consider whether the charge fluctuations can help grain grow
against the Coulomb potential barrier. The ratio of the grain kinetic energy
arising from charge
fluctuations to electrostatic potential at a distance equal to $2a$ with $a$
being the grain size is
\bea
\frac{E_{\kin}}{E_{\rm Coul}}=\frac{3a k_{\B}T_{d}^{\infty}}{\langle Z\rangle^{2}e^{2}}.\label{ratio}
\ena

From Equations (\ref{zmean}), (\ref{vd_ppd}), and (\ref{ratio}) we obtain
\bea
\frac{E_{\kin}}{E_{\rm Coul}}\simeq 2.3\times10^{1}\epsilon^{2}\left(\frac{a}{10^{-6}~\cm}\right)^{-5}
\left(\frac{T_{\gas}}{100~ \K}\right)\nonumber\\
\times\left(\frac{10^{10}~\cm^{-3}}{n_{\gas}}\right)\left(\frac{10^{-4}}{x_{i}}\right),~~~\label{ekin_coul}
\ena
where the ionization fraction $x_{i}\ne 0$ was assumed to be sufficiently high so that the charge
fluctuations are important.

Grain growth occurs only if $E_{\kin}/E_{\rm Coul}\ge1$.  Thus,
from Equation (\ref{ekin_coul}) we can derive the range of grain size with
the grain growth. Using parameters for PPDs conditions, $T_{\gas}=300~\K$, $n_{\gas}=10^{10}~\cm^{-3}$ and $x_{i}=10^{-4}$ and taking $\epsilon=10^{-1}$, we found that very small grains with size $a\leq 6\times 10^{-7}~\cm$. Therefore, the
growth of larger grains can be induced by other processes (e.g., turbulence as suggested in Okuzumi 2009).

The presence of PAHs and nanoparticles in PPDs can arise from the fragmentation
and shattering of micron graphite and silicate grains (see e.g.,
Dullemond \& Dominik 2008). The level of MHD turbulence in PPDs is very
uncertain, but it is believed that the layers near the surface are highly
turbulent. As a result, large grains can be accelerated by the resonant
interactions of fast modes to super-Alfv\'{e}nic speed (Yan et al. 2004;
Hoang et al. 2011b). Thus, the grain shattering may be efficient that return
very small grains to PPDs.

{\it (b) Effects of PAH acceleration on magnetorotational instability}

PAHs play an important role in the ionization equilibrium in PPDs.
Bai (2011) found that PAHs can favor the accretion due to
magneto-rotational instability (MRI) by decreasing ambipolar diffusions
in outer layers of PPDs. In fact, if charged PAHs are more abundant
than electron ($n_{\rm PAH}> n_{e}$), then the ambipolar diffusion
is less important. However, since the electrical conductivity by charged
PAHs is more efficient than electron, the MRI becomes more efficient.
Suprathermal velocities of PAHs due to charge fluctuations in PPDs
can enhance the accretion rate of electron and ion onto grains, which
certainly affect the ambipolar diffusion\footnote{The ambipolar diffusion is usually
invoked to describe the removal of magnetic field during star formation.
Another powerful mechanism is reconnection diffusion based on the
ability of turbulent magnetic field to reconnect and redistribute the entrained
matter (Lazarian 2005; see also Lazarian 2011).} and MRI.
Thus, understanding grain motions in PPDs is important for understanding
the dynamics of PPDs and planetesimal formation.

\section{Summary}

The present paper applies the Monte Carlo simulations to
quantify the efficiency of the new acceleration mechanism
for very small grains induced by charge fluctuations.
Our principal results can be summarized as follows.

1. Grain charge fluctuations due to both the sticking collisions with electrons
and ions in the plasma and the emission of photoelectrons by UV photons
are simulated using the Monte Carlo method. The charge distribution
obtained by MC simulations is similar to the steady charge distribution
for large grains, but they are different for very small grains.

2. Grain acceleration arising from electrostatic interactions between
dust grains is investigated using the Monte Carlo simulations.
We found that the acceleration induced by charge fluctuations is efficient
for very small grains, which can increase grain velocities to several times
their thermal velocities. This mechanism is more efficient
for the CNM with lower ionization fraction (i.e., larger Debye length) and larger charge dispersion
than the WNM and WIM.

3. The increase of relative velocities of PAHs and nanoparticles due
to charge fluctuations that we reported has certain effects
on grain coagulation and shattering and should be accounted for in
the models of grain evolution and planetesimal formation.

\acknowledgments
AL and TH acknowledge the support of the NSF-funded Center for Magnetic 
Self-Organization (CMSO) and the NASA grant NNX11AD32G. We thank Bruce Draine
for providing us the data of grain charge distribution and for fruitful discussions
and comments. We thank Hideko Nomura
for providing us the physical parameters of protoplanetary disks. We thank
Huirong Yan and Satoshi Okuzumi for valuable comments. We thank
the anonymous referee
for helpful comments and suggestions that improve the paper. AL acknowledges
Humboldt Award at the Universities of Bochum and Cologne and Visiting
Fellowship at the International Institute of Physics (Brazil).

\appendix
\section{A. Grain charging}
In plasma, grains can acquire charge due to sticking collisions with electron
and ions (see Draine \& Sutin 1987, DS87), and emissions of photoelectrons
due to UV photons.

\subsection{A1. Collisional charging}
First, let us briefly describe the collisional charging.
Consider an beam of charge particles $p$ with velocity $v$ colliding with a grain 
of charge $Q=Ze$, the
number of particles sticking to the grain per second is defined as the collisional charging rate:
\bea
{J}_{p,v}=s_{p}n_{p}v\sigma_{coll}=s_{p}n_{p} v \pi b_{\rm cr}^{2},
\ena
where $s_{p}$ is the sticking coefficient, $n_{p}$ is the number density of particles,
and $b_{\rm cri}$ is the critical impact factor such that the closest distance
between the particle and the charged grain is $r=a$. If $Z_{p}Z<0$, then
we easily get $b_{\rm cri}^{2}=\left(1-\frac{2Z_{p}Z}{mv^{2}a}\right)$ for all values of velocity $v$.
For $Z_{p}Z>0$, we have $b_{\rm cri}=\left(1-\frac{2Z_{p}Z}{mv^{2}a}\right)$
for velocity $mv^{2}/2>mv_{\rm cr}^{2}/2\equiv Z_{p}Ze^{2}/a$, i.e., kinetic energy
exceeds the repulsive Coulomb energy.

Assuming that particles have a Maxwellian distribution, we can obtain the charging rate
\bea
J_{p}(Z)&=&\int s_{p}n_{p}v\sigma_{coll}f_{\rm MW}(v) 4\pi v^{2} dv,
\ena
where $f_{\rm MW}(v)=\left(k_{\B}T_{\gas}/2\pi\right)^{-3/2}
\exp\left(-mv^{2}/{2k_{\B}T_{gas}}\right)$ is the Maxwellian 
distribution, and the integral limit is from $0-\infty$ for $Z_{p}<0$ and 
from $v_{\rm cri}-\infty$ for $Z_{p}Z>0$.

As a result, we obtain the charging rate from electron and ion sticking collisions:
\bea
J_{ion}=s_{i}n_{i}\left(\frac{8k_{\B}T_{i}}{\pi m_{i}}\right)^{1/2}\pi a^{2}\left[1-\left(\frac{Ze^{2}}{akT_{i}}\right)\right],\label{eq:Jion}\\
J_{e}=s_{e}n_{e}\left(\frac{8k_{\B}T_{e}}{\pi m_{e}}\right)^{1/2}\pi a^{2}{\rm \exp}\left(\frac{Ze^{2}}{akT_{e}}\right),\label{eq:Je}\\
\ena
for $Z<0$, and 
\bea
J_{ion}=s_{i}n_{i}\left(\frac{8k_{\B}T_{i}}{\pi m_{i}}\right)^{1/2}\pi a^{2}
{\exp}\left(\frac{Ze^{2}}{ak_{\B}T_{i}}\right),\label{eq:Jion1}\\
J_{e}=s_{e}n_{e}\left(\frac{8k_{\B}T_{e}}{\pi m_{e}}\right)^{1/2}\pi a^{2}\left[1+\left(\frac{Ze^{2}}{akT_{e}}\right)\right],\label{eq:Je1}\\
\ena
for $Z \ge 0$. Here $s_{i}$ is ion sticking coefficient, which is taken to
be unity (see DS87), $T_{i}$ and $T_{e}$ are ion temperature and electron
temperature. For isothermal plasma, $T_{i}=T_{e}=T_{\gas}$. Above we ignore the minor contribution from the
interaction of charge with the image potential
on the grain. Detailed descriptions can be found in given in DS87.

\subsection{A2. Photoemission}

The number of electrons leaving the grain due to photoemission by UV
photons per second is
\bea
J_{pe}(Z)=\pi a^{2}\int Y_{e}Q_{\rm abs}\frac{u_{\nu}c}{h\nu}d\nu,
\ena
where $n_{\nu}=u_{\nu}/(h\nu)$ is the density of photon with energy 
$E=h\nu> 13.6 eV$, and
$Y_{e}$ is the photoemission yield, and $Q_{\rm abs}$ is the absorption
efficiency (see detail in Weingartner \& Draine 2001).
For the ISM, we calculate photoemission rate $J_{pe}$ using the interstellar
radiation from Mathis et al. (1983). We adopt the tabulated data $Q_{\rm abs}$
for graphite grains from Draine \& Li (2001).

In the ISM, the photoemission is dominant, but in dense regions with high
density (e.g., protoplanetary disks) the collisional charging is dominant in
mid-plane regions where UV photons from the new born stars are shielded.
Close to the young stars or higher from the mid-plane,
the photoemission can be dominant due to abundance of UV photons.

\section{B. Gas Drag}

Interactions of dust grains with the ambient gas represent the primary mechanism of
dissipating streaming motions of grains. The
damping rate of translational motion arising from the interaction with neutral
gas is essentially the inverse time for collisions with the mass of the gas
equal that of a grain (Purcell 1969),
\bea
\gamma_{n}\equiv \tau_{dn}^{-1}
= 2\sqrt{\frac{2}{\pi}}\frac{n_n}{a\rho_d}(m_{n}k_{\B}T_n)^{1/2}
=2.4\times10^{-12}\left(\frac{10^{-6}\cm}{a}\right)\left(\frac{n_{n}}{30~\cm^{-3}}\right)
\left(\frac{m_{n}}{m_{\H}}\right)\left(\frac{T_{n}}{100~\K}\right)^{1/2},
\label{tau_fr}
\ena
where $m_n$, $n_n$, and $T_n$ are the mass, volume density, and temperature
of neutral atoms, $\rho_d$ is the mass density of dust grains, and $a$ is the
grain size.

When the ionization is sufficiently high, the interaction of charged grains with
ions in plasma becomes important (Draine \& Salpeter 1979). The ion-grain
cross section due to long-range Coulomb force is larger than the atom-grain
cross section. As a result, the rate of translational motion damping gets
modified (Draine \& Salpeter 1979).\footnote{This drag consists of both close
collisions and Coulomb distant interactions of dust with ions and electrons. Since the momentum
of electrons is much smaller than that of ions, drag due to collisions with electrons are disregarded.} 

For subsonic motions total damping rate due to neutral and ions is
\bea
\tau_{\drag}^{-1}\equiv \gamma_{dg}=\gamma_{n}+\gamma_{ion}=\alpha\gamma_{n},\label{eq:tau_drag}
\ena
where the renormalizing factor reads
\begin{eqnarray}
\alpha=1+\frac{n_{\rm H}}{2n_n}\sum_{i}x_i\left(\frac{e^{2}}{ak_{\B}T_i}
\right)^{2}\left(\frac{m_{i}}{m_n}\right)^{1/2}
\sum_{Z}Z^2f(Z)\nonumber\\
\times\ln\frac{3}{2\sqrt{\pi}}\frac{(k_{\B}T_i)^{3/2}}{|Z|e^3
(xn_{\rm H})^{1/2}}\hspace{.5cm}.\label{alpha1}
\end{eqnarray}
Here $x_{i}$ is the abundance of ion $i$ (relative to hydrogen) with mass
$m_{i}$ and temperature $T_i$, $x=\sum_i x_i$, and
$f(Z)$ is the grain charge distribution function. When the grain velocity $v_d$
relative to the gas becomes supersonic the dust interactions with the plasma is
diminished, and the damping rate in this
case is renormalized due to the gas-dynamic correction (Baines et al. 1965; Purcell 1969;
Draine \& Salpeter 1979) 
\bea
\alpha=\left(1+\frac{9\pi}{128}\frac{v_d^2}{C_{\rm s}^2}\right)^{1/2},\label{alpha2}
\ena
where $C_{\rm s}=\sqrt{k_{\B}T_{n}/m_n}$ is the sound speed.

\section{C. Theoretical Treatment for Acceleration due to Charge Fluctuations}

\subsection{C1. Charge evolution for large grains}
In Section 2 and 3, we discussed the charging and charge fluctuations for very small grains
and present Monte Carlo simulations to model the charge fluctuations. Here,
we consider the case of large grains for which the charge fluctuations can
be modeled using Langevin equations.

Assume that some small deviation from the mean grain charge is introduced, then
the charging rate changes accordingly to bring the charge to an equilibrium
state.
Denote $\tau_{\rm ch}$ be the relaxation time for which the charge fluctuations return
to the equilibrium state, the charging frequency becomes
\bea
\nu_{\rm ch}\equiv \tau_{\rm ch}^{-1}=\sum_{k}\frac{dJ_{k}}{dQ}_{Q=Q_{0}},\label{eq:nuch}
\ena
where $J_{k}$ with $k$=ion, e and pe, is the charging rate from the $k$ process.

For the collisional charging, using Equations (\ref{eq:Jion}) and (\ref{eq:Je}) for (\ref{eq:nuch}),
we obtain $\nu_{\ch}$ and the charge dispersion $\sigma_{Q}$ for a grain of size $a$
\bea
\nu_{\ch}=\frac{d(J_{ion}-J_{e})}{dQ}=\frac{1+z}{2\pi}\frac{a}{\lambda_{Di}}\omega_{i},\label{nuch}\\
\sigma_{Q}^{2}=\frac{1+z}{z(2+z)}|eQ_{0}|,\label{sigma2}
\ena
where $z=Q_{0}e/(akT_{i})$ and
\bea
\lambda_{{\rm D} i}=\frac{v_{T_i}}{\omega_{{\rm p}i}}=
\left(\frac{k_{\B}T_{i}}{m_{i}}\right)^{1/2}\left(\frac{m_{i}}{4\pi e^{2}n_{i}}
\right)^{1/2}=\left(\frac{k_{\B}T_{i}}{4\pi e^{2}n_{i}}\right)^{1/2}
\approx2\times 10^{2}\left(\frac{T_{i}}{100 {~\rm
 K}}\right)^{1/2}\left(\frac{0.1~\cm^{-3}}{n_{i}}\right)^{1/2},\label{eq1}
\ena
is the Debye length due to ion with the plasma frequencies
\bea
\omega_{pi}^{2}=\frac{4\pi e^{2}n_{i}}{m_{i}},~~\omega_{pd}^{2}=\frac{4\pi
Q_{0}^{2}n_{d}}{m_{d}}.\label{omega_p}
\ena

\subsection{C2. Continuous Approximation}

The charge fluctuations of large grains can be described by a Gaussian
and Markovian random process.
The time evolution of charge  can be modeled by a stochastic differential equation:
\bea
\frac{dQ}{dt}=-\nu_{ch}Q+\Gamma_{Q},
\ena
where
\bea
\langle \Gamma(t)\Gamma(t+\tau)\rangle=2\nu_{ch}\sigma_{Q}\delta(\tau).
\ena
where $\sigma_{Q}$ is the standard deviation of charge from equilibrium or
charge dispersion (see a review by Morfill \& Ivlev 2009).
The auto-correlation function of charge is a decaying law
\bea
\langle Q(t)Q(t+\tau)\rangle=Q_{0}^{2}exp(-\nu_{\ch}\tau).
\ena

\subsection{C3. Evolution of grain kinetic temperature}

In order to understand the effect of the energy variation in binary dust-dust 
collisions  on the
mean kinetic energy of the whole dust ensemble, one should employ the kinetic
approach. The kinetics is described in terms of
the velocity distribution function $f_d({\bf p},t)$. There are two principal
contributions to the dust kinetics -- one is
due to the mutual dust collisions and another because of dust interactions with
the ambient gas.

Three known interaction processes between dust-gas are dust-neutral, dust-
ion collisions, and plasma drag, which arises from interaction of passing
ions with grain dipole moment. We assume that these dust-gas interactions
lead to the local thermal equilibrium of gas and dust, characterized by a
temperature $T_g$, at different rates $\gamma_{n}, \gamma_{i}$
and $\gamma_{p}$ (see Appendix B).

The resulting kinetic equation has the following form:
\bea\label{01}
\frac{df_{d}}{dt}={\rm St}_{dd}f_{d}+{\rm St}_{dg}f_d,
\ena
where St$_{dd}$ and St$_{dg}$ denote the collision operators (integrals)
describing the dust-dust and dust-gas interactions,
respectively. Although the analysis of Eq. (\ref{01}) can be performed for
arbitrary $f_d({\bf p})$, for the sake of
convenience we assume that dust particles have a Maxwellian velocity distribution
$f_{\rm M}({\bf p})$. Then the equation
for evolution of mean kinetic energy (kinetic temperature) $kT_d(t)=
\frac13\int(p^2/m_d)f_{\rm M}d{\bf p}$ is
obtained by taking the second moment of Eq. (\ref{01}),

\bea\label{4a}
\dot T_d=\frac13\int\frac{p^2}{m_d}\left({\rm St}_{dd}f_{\rm M}+{\rm St}_{dg}
f_{\rm M}\right)d{\bf p}.
\ena

Ivlev et al. (2010) calculated integral (\ref{4a}) and derived the evolution
equation for $T_{d}$ as followings.

\bea\label{0}
\dot T_{d}=\frac{\sigma_{Q}^2\omega_{{\rm p}d}^2}{Q_0^2\nu_{\rm ch}}T_{d}
-2\gamma_{dg}(T_d-T_g),\label{dTdt_C}
\ena
where $\omega_{{\rm p}d}=\sqrt{4\pi Q_0^2n_d/m_d}$ is the dust plasma frequency,
and $\gamma_{dg}$ is the total damping rate due to dust-gas interactions.

From equation \ref{dTdt_C}, let denote the excitation coefficient
\bea
\alpha_{C}=\frac{\sigma_{Q}^2\omega_{{\rm p}d}^2}{Q_0^2\nu_{\rm ch}}.
\ena

Using typical parameters, we obtain the ratio of source-to-sink 
from equation (\ref{dTdt_C}):
\bea
\frac{\alpha_{C}}{\gamma_{dg}}=2\times 10^{-3} \left(\frac{\tilde{\sigma}_{Q}}{0.1}\right)^{2}
\left(\frac{Q_{0}}{100 ~e}\right)\left(\frac{n_{d}}{10^{3} \cm^{-3}}\right) \left(\frac{10^{2}}{\nu_{\ch}}\right)\left(\frac{1}{\gamma_{dg}}\right)\left(\frac{10^{-12} {\rm g}}{m_{d}}\right).
\ena
In order to have the increase of dust kinetic energy, the source term must exceed the sink term, i.e.,
$\alpha_{C}/\gamma_{dg}>1$.

In the low temperature regime, the dust-dust collision is similar to hard-sphere
collision. Thus, the variation of $T_{d}$ in time is governed by 
\bea
\dot T_{d}=\frac{\sigma_{Q}^2}{m_{d}Q_{0}^2\nu_{\rm ch}\lambda l}T_{d}^{2}
-2\gamma_{dg}(T_d-T_g),\label{dTdt_HS}
\ena
where $l=1/(n\sigma_{0})=1/(n_{d}\pi a_{HS}^{2})$ with
$a_{HS}=2\lambda \ln\left(Q_{0}^{2}/\lambda \epsilon_{r}\right)$, and $\lambda$ is the
screening length.

Without charge fluctuations, when the collisions between dust particles conserve
the energy, the equilibrium temperature of
dust grain is determined by interactions with the ambient gas, so that $T_d=T_g$. 
Random charge fluctuations provide an additional
energy source, and if the coefficient of the first (source) term in the r.h.s. 
of Eq. (\ref{0}) exceeds the damping rate
$2\tau_{dg}^{-1}$, then the dust temperature grows exponentially with time.

The kinetic coefficient ${\cal A}_{dd}$  is linearly 
proportional to the maximum scattering angle,
which we set equal to unity. Hence, we implicitly supposed that there are 
sufficiently small impact parameters $\rho$ that
ensure scattering at large angles, $\chi\gtrsim1$. However, since the lower 
bound of impact parameters is limited by the
particle radius, $\rho\gtrsim a$, this assumption is only valid if the kinetic 
temperature is below a certain critical value
$T_d^{\rm cr}$. From the relation $\chi\sim Q_0^2/\rho T_d$ (Lifshitz \& Pitaevskii 1981)
we readily deduce,
\begin{displaymath}
T_d^{\rm cr}\sim \frac{Q_0^2}{a}\equiv\frac{|Q_0|}{e}zT_i,
\end{displaymath}
so that the {\it actual} value of the maximum scattering angle is equal to 
$T_d^{\rm cr}/T_d$. Therefore, the exponential
temperature growth described by Eq. (\ref{0}) proceeds until $T_d$ reaches the 
critical value $T_d^{\rm cr}(\gg T_i)$. At
larger temperatures the source term in Eq. (\ref{0}) remains  constant
(which is obtained by replacing $T_d$ with
$T_d^{\rm cr}$). The temperature growth switches to linear and eventually
gets saturated due to the translational damping, which determines the ultimate temperature of dust.

Note that when treating dust-dust collision, we assume the dust velocity
does not change as a result of gas collisions, which corresponds to Orbital-Motion-Limited
approximation (OML).

\subsection{C4. Analytical estimate for saturated kinetic temperature}

When $\alpha_{C}>\gamma_{dg}$, Equation (\ref{dTdt_C}) indicates that
the temperature $T_{d}$ increases exponentially with time. Thus, we call this
regime {\it instability}. When $T_{d}$ becomes larger
than the critical temperature $T_{d}^{cr}$ -limit of Coulomb interaction,
then the source term $\alpha_{C}$ is constant. As $T_{d}$ continues to increase,
the sink term increases while the source term remains constant. Therefore,
the temperature is saturated. From Equation (\ref{dTdt_C}), we obtain
\bea
T_{d}^{\infty}=\frac{\sigma_{Q}^{2}\omega_{pd}^{2}}{Q^{2}\nu_{ch}}
\frac{\tau_{dn}}{2\alpha}T_{d}^{cr}.\label{Td_sat}
\ena

Plugging $T_{d}^{cr}$ and  parameters $\nu_{ch}, \sigma_{Q}$ and 
$\tau_{dn}$ in eq. (\ref{Td_sat}) it yields
\bea
\frac{T_{d}^{\infty}}{T_{i}}=4\pi\frac{n_{i}}{n_{n}}
\frac{\lambda_{Di}^{4}n_{d}(a)}{a}\frac{z^{2}}{0.3(2+z)},\label{Tterm}
\ena
where $z=Q_{0}e/(akT_{i})$ has been used.

Using the usual expression for thermal velocity
\bea
v_{d}^{\infty}=\left(\frac{3k_{\B}T_{d}}{m_{d}}\right)^{1/2},
\ena
and assuming  the MRN dust size
distribution, we obtain
\bea
v_{T_d}^{\infty}\approx\frac{6\times 10^{2}}{(\alpha n_i)^{1/2}}\left
(\frac{T_{i}}{100{~\K}}\right)^{3/2}
\left(\frac{a}{10^{-6}{~\cm}}\right)^{-3.25}{\cm~\s^{-1}},\label{vd}
\ena
where $n_i$ is in cm$^{-3}$ and the renormalizing factor $\alpha$ for 
the damping rate in the subsonic and supersonic
regimes is given by Eq. (\ref{alpha1}) and (\ref{alpha2}), respectively. Note
that in the latter case $\alpha$ is the
function of dust velocity and hence Eq. (\ref{vd}) should be resolved for
$v_{T_d}^{\infty}$.

When the source term is smaller than the sink term, $\alpha_{C}< \gamma_{dg}$,
the dust temperature increases with time until the equilibrium with gas is 
established, and no instability exists. From Equation (\ref{dTdt_C}) 
we obtain
\bea
T_{d}^{sat}=\frac{1 }{1-\alpha_{C}/\gamma_{dn}} T_{n}.
\ena 
As $\sigma_{Q}$ increases, $T_{d}^{sat}$ increases with $\sigma_{Q}$


\begin{thebibliography}{}

\bibitem[]{} Acke, B.,\& van den Ancker, M. E. 2004, A\&A, 426, 151
\bibitem[]{} Baines, M.~J. 1965, \mnras, 130, 63
\bibitem[]{} Bai, X. N. 2011, \apj, 739, 51
\bibitem[Cho \& Lazarian(2003)]{2003MNRAS.345..325C} Cho, J., \& Lazarian, 
A. 2003, \mnras, 345, 325

\bibitem[Cho \& Lazarian(2002)]{2002PhRvL..88x5001C} Cho, J., \& Lazarian, 
A. 2002, Physical Review Letters, 88, 245001
\bibitem[]{}Chokshi, A., Tielens, A. G. G. M., Hollenbach, D. 1993, \apj, 407, 806
\bibitem{CG94} Cui, C., \& Goree, J. 1994, IEEE Trans. Plasma Sci. 22, 151

\bibitem[Draine(1985)]{1985prpl.conf..621D} Draine, B. T. 1985, Protostars and Planets II (Tucson, AZ: Univ. Arizona Press),
621

\bibitem[Draine(2003)]{2003ARA&A..41..241D} Draine, B. T. 2003, \araa, 41, 241

\bibitem[]{} Draine, B. T. 2009, in ASP Conf. Ser. 414, Cosmic Dust—Near and Far, ed. Th.
Henning, E. Grun, \& J. Steinacker (San Francisco, CA: ASP), 453

\bibitem[]{} Draine, B. T. 2011, Physics of the Interstellar and Intergalactic
Medium (Princeton, NJ: Princeton Univ. Press)

\bibitem[]{} Draine, B.~T., \& Lazarian, A. 1998, \apj, 508, 157
\bibitem[]{Draine+Li_2007} Draine, B. T., \& Li, A. 2007, \apj, 657, 810
\bibitem[]{DrSu87} Draine, B. T., \& Sutin, B. 1987, \apj, 320, 803 (DS87)

\bibitem[Draine \& Salpeter(1979)]{1979ApJ...231..438D} Draine, B. T., \& 
Salpeter, E. E. 1979, \apj, 231, 438

\bibitem[]{} Dullemond CP, \& Dominik C. 2008, A\&A, 487



\bibitem[]{}Fraisse, A. A., et al. 2009, in AIP Conf. Ser. 1141, CMB Polarization Workshop: Theory and Foregrounds: CMBPol Mission Concept Study, ed. S. Dodelson,
et al. (Melville, NY: AIP), 265

\bibitem[]{}Geers, V. C., et al. 2006, A\&A, 459, 545


\bibitem{HY09} Hirashita, H., \& Yan, H. 2009, \mnras, 394, 1061

\bibitem[]{Hoang10} Hoang, T., Draine, B.~T., \& Lazarian, A. 2010, \apj, 715, 1462
\bibitem[]{}Hoang, T., Lazarian, A., \& Draine, B. T. 2011a, \apj, 741, 87
\bibitem[]{}Hoang, T., Lazarian, A., \& Schlickeiser, R. 2011b, \apj, in press (arXiv: 1111.4024)
\bibitem{Ivlev04} Ivlev, A. V., Zhdanov, S. K., Klumov, B. A., Tsytovich, V. N.,
 de Angelis, U., \& Morfill, G. E. 2004, Phys. Rev. E, 70, 066401

\bibitem{Ivlev10} Ivlev, A.V., Lazarian, A., Tsytovich, V.N., 
de Angelis, U., Hoang, T., \& Morfill, G. E. 2010, \apj, 723, 612

\bibitem{Ivlev05} Ivlev, A. V., Zhdanov, S. K., Klumov, B. A., \& Morfill, G. 
E. 2005, Phys. Plasmas 12, 092104

\bibitem[]{} Lazarian, A. 2007, J. Quant. Spectrosc. Radiat. Transfer, 106, 225
\bibitem[Lazarian \& Finkbeiner(2003)]{2003NewAR..47.1107L} Lazarian, A., \&
Finkbeiner, D. 2003, New Astronomy Review, 47, 1107
\bibitem[Lazarian \& Yan(2002)]{2002ApJ...566L.105L} Lazarian, A., \& Yan, 
H. 2002, \apjl, 566, L105

\bibitem{Landau_Kinetics} Lifshitz, E. M., \& Pitaevskii, L. P. 1981, {\it Physical 
Kinetics} (Oxford: Pergamon)
\bibitem[]{Mathis, Rumpl, Nordsieck 1977} Mathis, J. S.,  Rumpl, W., \& 
Nordsieck, K. H. 1977, \apj, 217, 425
\bibitem[]{767} Mathis, J., Mezger, P., \& Panagia, N. 1983, A\&A, 128, 212

\bibitem{Okuzumi09} Okuzumi, S. 2009, \apj, 698, 1120

\bibitem[]{} Oliveira, I., Pontoppidan, K. M., Mer\'{i}n, B., van Dishoeck, E. F., Lahuis, F., et al. 2010., \apj, 714, 778
\bibitem[]{} Ormel, C. W., Paszun, D., Dominik, C., \& Tielens, A. G. G. M. 2009, A\&A, 502, 845
\bibitem[]{} Ossenkopf, V. 1993, A\&A, 280, 617
\bibitem[]{} Perez-Becker, D., \& Chiang, E. 2011, \apj, 735, 8 
\bibitem[Purcell(1969)]{1969Phy....41..100P} Purcell, E. M. 1969, Physica, 41, 100

\bibitem[]{}Tanaka, H., Himeno, Y., \& Ida, S. 2005, \apj, 625, 414
\bibitem[Tielens(2005)]{2005pcim.book.....T} Tielens, A.~G.~G.~M. (ed.) 2005, in The Physics and Chemistry of the Interstellar Medium (Cambridge:Cambridge Univ. Press)


\bibitem{Spitzer} Spitzer, L. 1978 {Physical Processes in the interstellar medium}
 (New York)
\bibitem{VanKampen} van Kampen, N. G. 1981, {Stochastic Processes in Physics and 
Chemistry} (Amsterdam: Elsevier)

\bibitem[]{} van Boekel, R., Waters, L. B. F. M., Dominik, C., Bouwman, J., de Koter, A., Dullemond, C. P.,\& Paresce, F. 2003, A\&A, 300, L21
\bibitem{Yan04} Yan, H., Lazarian, A., \& Draine, B. T. 2004, ApJ, 616, 895

\bibitem[Yan \& Lazarian(2003)]{2003ApJ...592L..33Y} Yan, H., \& Lazarian, A.
 2003, \apjl, 592, L33
\bibitem[Yan]{} Yan, H., \& Lazarian, A. 2008, \apj, 673, 942
\bibitem[Yan]{} Yan, H. 2009, \mnras, 397, 1093
\bibitem[Weingartner \& Draine(2001)]{2001ApJ...563..842W} Weingartner, J. C.,
\& Draine, B. T. 2001, \apj, 563, 842
\bibitem[]{} Weingartner, J., \& Draine, B.~T. 2001, \apjs, 134, 264
\bibitem[]{} Whittet, D. C. B. 2003, Dust in the Galactic Environment (Bristol: Institute of Physics Publishing)
\end{thebibliography}
\end{document}